# Ground state and Spin-Wave dynamics in Brownmillerite SrCoO$_{2.5}$ – A combined Hybrid Functional and LSDA+$U$ study

Chandrima Mitra, Randy S. Fishman, Satoshi Okamoto, Ho Nyung Lee and Fernando A. Reboredo

Materials Science and Technology Division, Oak Ridge National Laboratory, Oak Ridge, Tennesse, 37831, USA

**Abstract**

We theoretically investigate the ground state magnetic properties of the brownmillerite phase of SrCoO$_{2.5}$. Strong correlations within Co $d$ electrons are treated within the local spin density approximations of Density Functional theory (DFT) with Hubbard U corrections (LSDA+$U$) and results are compared with the Heyd Scuzeria Ernzerhof (HSE) functional. The parameters computed with a $U$ value of 7.5 eV are found to match closely to those computed within the HSE functional. A *G*-type antiferromagnetic structure is found to be the most stable one, consistent with experimental observation. By mapping the total energies of different magnetic configurations onto a Heisenberg Hamiltonian we compute the magnetic exchange interaction parameters, J, between the nearest neighbor Co atoms. The J's obtained are then used to compute the spin-wave frequencies and inelastic neutron scattering intensities. Among four spin-wave branches, the lowest energy mode was found to have the largest scattering intensity at the magnetic zone center, while the other modes becomes dominant at different momenta. These predictions can be tested by experimentally.

## I. Introduction

The diverse physical properties exhibited by perovskite oxides ($ABO_3$) make them an important class of materials. They range from ferroelectricity [1] and magnetism [2] to colossal magnetoresistance and superconductivity [3]. This interplay between structural, magnetic and transport properties makes perovskites particularly interesting. Among the perovskites, cobalt based materials are becoming increasingly popular as a replacement for platinum in catalytic converters in diesel vehicles [4]. Moreover, the ones with oxygen vacancies, have been reported to show high oxygen mobility which make them promising candidates for use in catalysis, gas sensing, or as an oxygen membrane in solid oxygen fuel cells [5]. These oxygen deficient perovskites ($ABO_{3-\delta}$) show subtle changes in their physical properties compared to their bulk counterparts [6,7]. $SrCoO_{3-\delta}$ is one such example. While stoichiometric $SrCoO_3$ has a cubic perovskite structure, its crystal structure is considerably modified depending on the amount of oxygen vacancies in it [8,9,10]. More specifically the oxygen deficient $SrCoO_{3-\delta}$ shows a wide variety of phase changes when $\delta$ varies from 0.25 to – 0.5 [8,9,10].

$SrCoO_{2.5}$ is particularly interesting because the oxygen vacancies form an ordered arrangement. The brownmillerite-type structure is formed by quenching this oxygen deficient compound in liquid N$_2$ [11], which orders the oxygen vacancies. The crystal structure of $SrCoO_{2.5}$ is composed of alternating $CoO_6$ octahedron and $CoO_4$ tetrahedron layers along the x or the c axis depending on the choice of convention [12,13]. It can be derived from the perovskite structure by ordering oxygen vacancies along [011] direction. Indeed, if additional oxygen atoms are forced into these deficient positions under high pressure, the perovskite structure is subsequently restored [8,11]. In a recent experimental study by Choi et. al. [25], the reversal of the lattice and electronic structure evolution in SrCoO$_x$ (x=2.5-3) was directly observed using real-time spectroscopic ellipsometry. They showed that the two phases (SrCoO$_{2.5}$ and SrCoO$_3$) "could be reversibly controlled by changing the ambient pressure at greatly reduced temperatures". While the stoichiometric perovskite, SrCoO$_3$ is a ferromagnet and a metal, experimental observations indicate that brownmillerite phase, SrCoO$_{2.5}$ is antiferromagnetic and insulating with a Neel temperature of about 570 K [14].

The goal of this paper is to study the electronic structure and ground state magnetic properties of SrCoO$_{2.5}$. As is well known, within DFT, commonly used local and semi-local approximations to the exchange correlation functional, such as the Local Density Approximation (LDA) and Generalized Gradient Approximation (GGA), fail to produce an accurate description of the electronic structure. In particular the computed band gaps of semiconductors and insulators are underestimated within these approximations. Hence one needs to go beyond LDA or GGA in order to improve on this shortcoming. In order to address this issue we will employ two different approaches to compute the electronic structure of SrCoO$_{2.5}$. As will be discussed in detail in the following section, the first is the so called Local Spin Density +$U$ (LSDA+$U$) approximation [ref] which applies a Hubbard $U$ like "correction" to the strongly correlated orbitals, which in our case are the *Co d* orbitals. However, although computationally cheap, $U$ remains an empirical parameter here. A different route is the application of Hybrid Functionals which corrects for self-interaction errors. A comparative study of these two functionals would then allow us to obtain an "optimal" value of the $U$ parameter in the LSDA+U calculations.

In order to compute the magnetic exchange interaction parameters between the *Co* atoms we rely on a mean field Heisenberg-Hamiltonian. A Heisenberg model with only nearest neighbor interactions will be used to fit the interaction parameters J to the total energies obtained within DFT. Henceforth, *Co* atoms lying on the octahedral plane will be referred to as *Co$_{oct}$*, while those lying on the tetrahedral plane would be denoted as *Co$_{tet}$*. We find that all the interaction parameters J are negative. They include couplings within the octahedral and the tetrahedral plane as well as couplings between the two planes. The ground state magnetic configuration is hence found to be G-type, in agreement to experimental findings [14]. From the computed values of the interaction parameters, we predict the spin dynamics of the system using a 1/S expansion starting from the classical ground state [15,16].

The rest of the paper is organized in the following manner. Sec. II provides the computational details, Sec. III presents the results for the magnetic interaction parameters, in Sec. IV the Spin density wave formalism and results are discussed and finally we draw our conclusions in Sec. V.

## II. Computational method

The electronic structure calculations have been performed within DFT employing the Vienna Ab-initio Simulation Package (VASP) code [17]. The projector augmented wave pseudopotentials (PAW) have been used [18]. Valence electrons included for Sr, Co and O are $4s^2\,4p^6\,5s^2$, $3s^2\,3p^6\,4s^2\,3d^2$ and $2s^2\,2p^4$ respectively and a plane wave cut-off energy of 600 eV has been used. A Monkhorst-Pack special k-point grid [19] of 4×6×6 was chosen to integrate over the Brillouin zone. The energies are converged to within $10^{-6}$ eV/cell. All forces are converged within 0.004 eV/Å.

For the LSDA+$U$ calculations we used the simplified or the rotationally invariant approach as introduced by Dudarev *et al.* [20] and is implemented in the VASP code. Within this approach the LSDA+$U$ functional is written as

$$E_{LSDA+U} = E_{LSDA} + \frac{U-J}{2}\sum_{\sigma} Tr\rho^{\sigma} - Tr(\rho^{\sigma}\rho^{\sigma}) \qquad (1)$$

where $\rho^{\sigma}$ is the density matrix of $d$ electrons. $U$ and $J$ are the spherically averaged matrix elements of the screened Coulomb electron-electron interaction. In order to determine an "optimal" value of the $U$ parameter we compared the results to a hybrid functional calculation which is known to improve the band gaps for many semiconductors and insulators. We specifically used the Heyd Scuzeria Ernzerhof (HSE) functional as implemented in VASP.

Within the HSE formalism [21] the exchange correlation functional is constructed from a fraction of Hartree-Fock exchange ($E_x$) and the generalized gradient approximation due to Perdew, Burke and Ernzerhof (PBE) [22]. The method has an advantage over the PBEh [23] hybrid functional due to its faster convergence. This is because in HSE, as proposed by Heyd. *et al.* [24], the exact nonlocal exchange is further decomposed into a long range and a short range part in real space. The range separation is determined by a parameter, μ, which is typically chosen as a distance at which the non-local long range interaction becomes negligible. The HSE exchange correlation functional is written as:

$$E_{xc}^{HSE} = \alpha E_x^{sr,\mu} + (1-\alpha)E_x^{PBE,sr,\mu} + E_x^{PBE,lr,\mu} + E_c^{PBE} \quad , \qquad (2)$$

where $\alpha$ is called the 'mixing parameter'; which accounts for the amount of Hartree-Fock like exchange in the exchange correlation functional. The superscript $sr$ and $lr$ denotes the short range and long range part, respectively and $\mu$ refers to the screening parameter as mentioned earlier. The HSE calculation yields a band gap of 0.6 eV. We find that this corresponds to an $U$ of 7.5 eV. We keep the $J$ parameter fixed to 1 eV. From an optical absorption spectra Lee *et. al.* [25] obtained a band gap of 0.35 eV at 300 K which opened up to 0.45 eV at 5K.

### III. Results and discussions

The orthorhombic unit cell of brownmillerite SrCoO$_{2.5}$ is shown in Fig. (1). Although the assignment of space group to SrCoO$_{2.5}$ has been controversial [12] due to the existence of both the *Ima2* and the *Pnma* structure, we have used the *Ima2* structure in all our calculations as this was predicted to be more stable by *ab-inito* calculations [12]. We have used the experimental lattice constants where $a = 15.735$ Å, $b = 5.572$ Å and $c = 5.466$ Å [12]. Although in a recent work [10] the c axis was chosen as the longer axis, this does not change the overall physics and hence would not affect the results that are presented in this work. The presence of ordered oxygen vacancy "channels" in SrCoO$_{2.5}$ modifies the crystal structure from its cubic perovskite phase. The *Co-O* bond length changes which results in tilting of the *Co-O* polyhedral as can be seen in Fig. 1. In Table 1 we present the *Co-O* bond distances computed within HSE and LSDA+$U$, where two different values of $U$ have been used. While the LSDA+$U$ bond lengths are slightly overestimated compared to experimental values, the HSE bond distances are slightly underestimated. However, overall there is reasonable agreement.

The spin polarized $d$ states of *Co$_{oct}$*, *Co$_{tet}$* and O $p$ states computed within HSE are presented Fig. 2(a), 2(b) and 2 (c) respectively. We find an insulating ground state for SrCoO$_{2.5}$. The corresponding density of states plot for an LSDA+$U$ calculation is presented in Fig. 3 where an $U$ value of 7.5 eV has been used to match a band gap of about 0.6eV within HSE. Co in SrCoO$_{2.5}$ has a valency of +3 and therefore should have 6 valence electrons. However, as can be seen from Fig. 2(a), (b) and (c), there is a strong mixing of the *Co d* and the oxygen *p* states. Taking into account this hybridization, we find that Co effectively has 7 electrons occupying its $d$ states instead of 6. Furthermore the five-fold d levels split due to the crystalline field which no longer has a perfect octahedral or tetrahedral symmetry due to the Co-oxygen polyhedral rotations as shown in Fig. (1). The formation of high spin state on both *Co$_{oct}$* and *Co$_{tet}$* indicates that the exchange splitting dominates over the crystalline field splitting between the $d$ levels.

## III A.  Magnetic exchange parameters

We now move on to compute the magnetic exchange parameters. We assume that a Heisenberg Hamiltonian model describes the interaction between localized moments on the *Co* atoms as

$$H = -\sum_{ij} J_{ij} \mathbf{S_i} \cdot \mathbf{S_j} + E_0 \quad (3)$$

where *J's* are the interaction parameters, **S** indicates the localized spin on each site and $E_0$ is an energy reference. The first neighbor interaction between $Co_{oct}$ atoms is referred as $J_1$; the interaction among the $Co_{tet}$ is referred to as $J_2$; finally, the inter-planar interaction between $Co_{oct}$ and $Co_{tet}$ is $J_3$ (Fig. 3). Positive values for *J's* indicate ferromagnetic and negative values mean antiferromagnetic interactions. We thus have four unknown parameters, $J_1$, $J_2$, $J_3$ and $E_0$ and hence would need four different configurations to determine them. The four different configurations that we used are shown in in Fig. 4(a), (b), (c) and (d), where 4(a) represents the ferromagnetic configuration (FM), 4(b),(c) and (d) represents the three different antiferromagnetic configurations, AFM (I), AFM(II) and AFM (III) respectively. The blue spheres represent the Co atoms that are octahedrally coordinated and the green spheres represent the ones that are tetrahedrally coordinated. The red spheres are the oxygen atoms. Sr atoms have not been included in the picture.

Considering only nearest neighbor interactions in the Hamiltonian (Eq. 3), one can therefore write down four sets of equations which represent these four equations as follows :

$$E_{FM} = E_0 - 4S^2 J_1 - 2S^2 J_2 - 2S^2 J_3 \quad (4)$$

$$E_{AFM(I)} = E_0 - 4S^2 J_1 - 2S^2 J_2 + 2S^2 J_3 \quad (5)$$

$$E_{AFM(II)} = E_0 - 4S^2 J_1 + 2S^2 J_2 \quad (6)$$

$$E_{AFM(III)} = E_0 + 4S^2 J_1 - 2S^2 J_2 \quad (7)$$

where **S** = 3/2 for all the above equations. Equation (4) and (5) represent the configurations presented in Fig 3 (a) and 3(b), respectively. The other two configurations can be similarly set up. It must be pointed out that the Co atoms, that are tetrahedrally coordinated, are assumed to have two nearest neighbors (instead of four, as is the case for octahedral coordination) due to the absence of two oxygen atoms. Upon solving equations 2,3,4 and 5 we obtain $J_1$, $J_2$ and $J_3$. We calculate the energies of each configuration with DFT. In Table II and III we present the total energies and the *J*'s, computed within both LSDA+*U* and HSE. As can be seen the *J*'s computed within HSE match closely to an *U* value of 7.5 eV.

All the *J*'s are found to be negative implying antiferromagnetic interaction between the *Co* atoms. We also find the intra-planar interactions, $J_1$ and $J_2$ to be stronger than the inter-planar interaction $J_3$. This confirms the experimental observation of $SrCoO_{2.5}$ being a *G*-type antiferromagnet. The *G*-type antiferromagnetic configuration is presented in Fig. 5 and is indeed found to have the lowest energy when compared to all the other configurations presented above. We further check the effect of supercell size on the magnetic interactions by employing a 1x2x2 supercell and find all our results converged to within 0.1 meV.

## I. Spin waves

A detailed formalism for computing the spin-wave frequencies and intensities have been discussed elsewhere [15, 16, 26]. In this paper we will be using the same notations as used in [15]. Following the Holstein-Primakoff transformation the spin operators $S_{i+}$ and $S_{i-}$ are transformed to boson creation and annihilation operators $a_i$ and $a_i^+$ as $S_{i+} = \sqrt{2S} a_i$ and $S_{i-} = \sqrt{2S} a_i^+$. Due to the spin structure of the *G*-type AF phase, the magnetic unit cell contains 8 sublattices. The Hamiltonian *H* is expanded in powers of $1/\sqrt{S}$ about the classical limit as $H = E_0 + H_1 + H_2 + \ldots$ In this expansion of the Hamiltonian, $E_0$ is the classical energy and $H_2$ describes the dynamics of the spin waves (SW). In equilibrium $H_1$ must vanish.

In order to determine the SW frequencies, $\omega^{(r)}(\mathbf{k})$, one needs to solve the equation :

$$id\mathbf{v_k}/dt = -[H_2, \mathbf{v_k}] = M(\mathbf{k})\mathbf{v_k} \qquad (8)$$

where $\mathbf{v_k} = [a_\mathbf{k}^{(1)}, a_\mathbf{k}^{(1)+}, a_\mathbf{k}^{(2)}, a_\mathbf{k}^{(2)+}, \ldots, a_\mathbf{k}^{(s_t)}, a_\mathbf{k}^{(s_t)+}]$ and $M(\mathbf{k})$ is written as a $2s_t \times 2s_t$ matrix, $s_t =$ 8 being the number of spin sublattices on four inequivalent layers. The SW frequencies, $\omega^{(r)}(\mathbf{k})$, are obtained from Det$[M(\mathbf{k})N - \omega^{(r)}(\mathbf{k})I/2] = 0$ where *N* is a diagonal matrix with upper 8 matrix elements +1 and lower 8 matrix elements -1 and *I* is the 16-dimensional unit matrix. The dispersion relations for $\omega^{(r)}(\mathbf{k})$ are plotted in Fig. 6(a). Each of the four magnon modes is doubly degenerate. A gap of about 10 meV separates the low-frequency acoustic and higher-frequency optical SW modes.

Due to the missing *Co*-Oxygen bonds along the (1,1,0) direction, the solid branch of acoustic modes is higher between (0,0,4) and (1,0,4) than between (0,0,4) and (0,1,4). While the solid branch vanishes at (0,0,4), (1,0,4), and (0,1,4), the dashed branch vanishes at (0,0,2).

We further determine the weight of each SW frequency from the eigenvectors of the matrix $M(\mathbf{k})$. The spin-spin correlation function $S_{ab}(\mathbf{k}, W)$ can be expressed as a sum over delta functions at each SW frequency [14] and the inelastic neutron-scattering intensity is given by

$$S(\mathbf{k},\omega) = \sum_{\alpha\beta}(\delta_{\alpha\beta} - \frac{k_{\alpha\beta}}{k^2})S_{\alpha\beta}(\mathbf{k},\omega) = \sum_r \partial(\omega - \omega^{(r)}(\mathbf{k}))A_r(\mathbf{k}) \quad (9)$$

The second term implies that the observed spin excitations are polarized transverse to the momentum $\mathbf{k}$.

For the solid branch of acoustic modes, the SW intensity $A_r(\mathbf{k})$ plotted in Fig.6(b) diverges at the ordering wavevectors (1,0,4) and (0,1,4). But $A_r(\mathbf{k})$ vanishes for the dashed acoustic mode at (0,0,2) and for the solid acoustic mode at (0,0,4), where those mode frequencies are zero. Whereas the dash-dot optical mode always has zero weight, the small dash optical mode with frequency between 100 and 130 meV dominates the SW spectrum near (0,0,4). This mode should also be readily observable near (1/2,1/2,4). Three modes may be observed between (0,1,4) and (0,0,4) close to (0,0,4).

## I. Conclusions

In conclusion we have performed a first-principles study of the electronic structure and magnetic properties of SrCoO$_{2.5}$, employing Hybrid Functional and an LSDA+$U$ approach. We find *Co* to be in high spin state. The ground state magnetic exchange interaction parameters were computed and the Holstein-Primakoff expansion was employed to predict the SW dispersion as well as the structure factor of this system. Our study confirms the *G*-type magnetic configuration where all the parameters are found to be negative. The intra-planar magnetic exchange between *Co* atoms are found to be stronger than the inter-planar exchange. The structure and magnetic properties of brownmillerite compounds and the effect of doing them with different magnetic elements have been a subject of study such as that in Ca$_2$Fe$_{1-x}$Mn$_x$AlO$_{5+\delta}$.

In future works it would be interesting to study the effect of doping brownmillerite SrCoO$_{2.5}$, with impurity atoms, on the valence and spin state of *Co* atoms. This in turn would affect the magnetic exchange between the *Co* atoms. A detailed understanding of this could enable us to 'design' different magnetic states within this oxygen deficient cobaltite.

**Acknowledgments**

This work was supported by the US Department of Energy, Basic Energy Sciences, Materials Sciences and Engineering Division.

**Figure Captions**

Fig. 1: (Color online) Crystal structure of $SrCoO_{2.5}$. The Cobalt, Oxygen and Strontium atoms are represented by blue, red and green spheres respectively.

Fig. 2 : Partial density of states computed within HSE (a) $d$ states of Co which are octahedrally coordinated ($Co_{oct}$). (b) $d$ states of Co which are tetrahedrally coordinated ($Co_{tet}$). (c) Oxygen $p$ states.

Fig. 3 : Partial density of states computed within LSDA+$U$ ($U$=7.5 eV) (a) $d$ states of Co which are octahedrally coordinated ($Co_{oct}$). (b) $d$ states of Co which are tetrahedrally coordinated ($Co_{tet}$). (c) Oxygen $p$ states.

Fig. 4 : (Color online) Different magnetic configurations used in calculating the exchange parameters, $J_1$, $J_2$ and $J_3$. The direction of spins have been indicated by the arrows. (a) FM. (b) AFM(I). (c) AFM(II). (d) AFM(III).

Fig. 5 : (Color Online) *G*-type Antiferromagnetic structure.

Fig. 6: (Color Online) (a) Predicted spin wave frequencies $\omega^{(r)}$. (b) Corresponding inelastic neutron scattering intensities, $A_r(\mathbf{k})$, as a function of momentum. Parameters used are the $J_1$, $J_2$ and $J_3$ computed within HSE, as presented in Table III.

**Tables**

Table I: Cobalt-Oxygen (*Co-O*) bond lengths as computed within HSE and different *U* parameters. $O_{oct}$, $O_{tet}$ and $O_{int}$ refer to the Oxygen atoms in the octahedral, tetrahedral and the intermediate layers, respectively, as labelled in Fig. 1.

|  | $Co_{oct}$-$O_{oct}$ (Å) | $Co_{tet}$-$O_{tet}$ (Å) | $Co_{oct}$-$O_{int}$ (Å) | $Co_{tet}$-$O_{int}$ (Å) |
|---|---|---|---|---|
| *U*=4 (eV) | 1.954 | 2.058 | 2.200 | 1.785 |
| *U*=7.5 (eV) | 1.970 | 2.060 | 2.231 | 1.820 |
| **HSE** | 1.949 | 2.025 | 2.200 | 1.790 |
| **Exp[12]** | 1.950 | 2.030 | 2.210 | 1.801 |

Table II: Total energies of different magnetic configurations computed with *HSE* and LSDA+*U* functionals.

|  | *HSE* | *U*=7.5 | *U*= 4 |
|---|---|---|---|
| $E_{FM}$ (eV) | -352.22 | -232.23 | -242.52 |
| $E_{AFM(I)}$ (eV) | -352.38 | -232.43 | -243.03 |
| $E_{AFM(II)}$ (eV) | -352.50 | -232.56 | -243.32 |
| $E_{AFM(III)}$ (eV) | -352.73 | -232.78 | -243.86 |

Table III: The magnetic exchange interaction parameters.

| | $J_1$ (meV) | $J_2$ (meV) | $J_3$ (meV) |
|---|---|---|---|
| $U = 4$ eV | -60 | -61 | -57 |
| $U = 7.5$ eV | -25 | -26 | -22 |
| *HSE* | -24 | -22 | -18 |

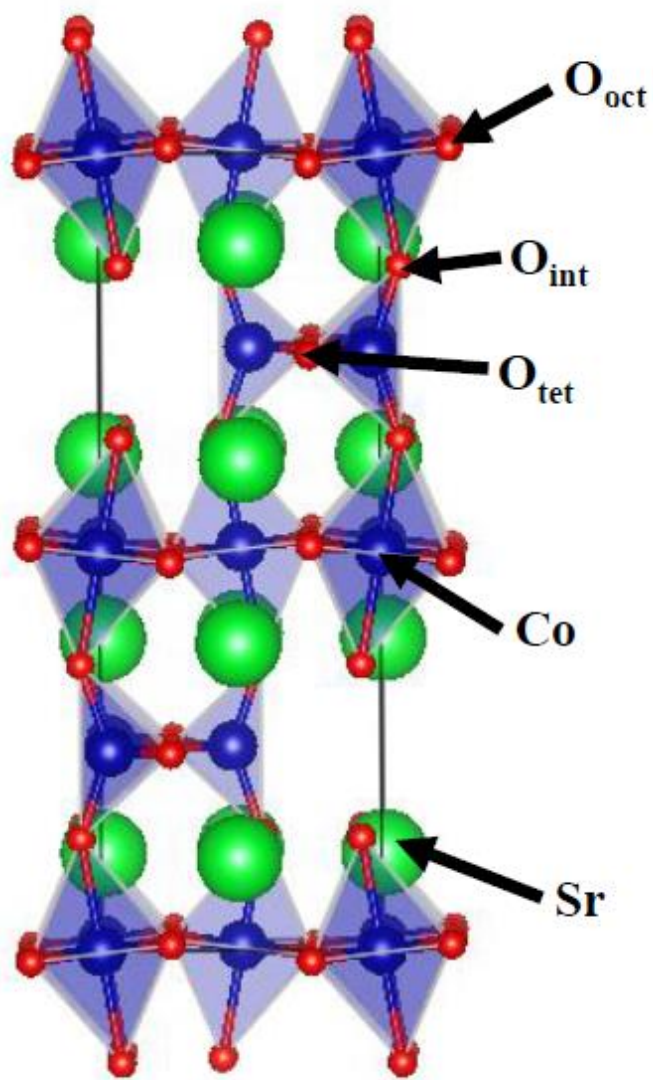

**Fig. 1**

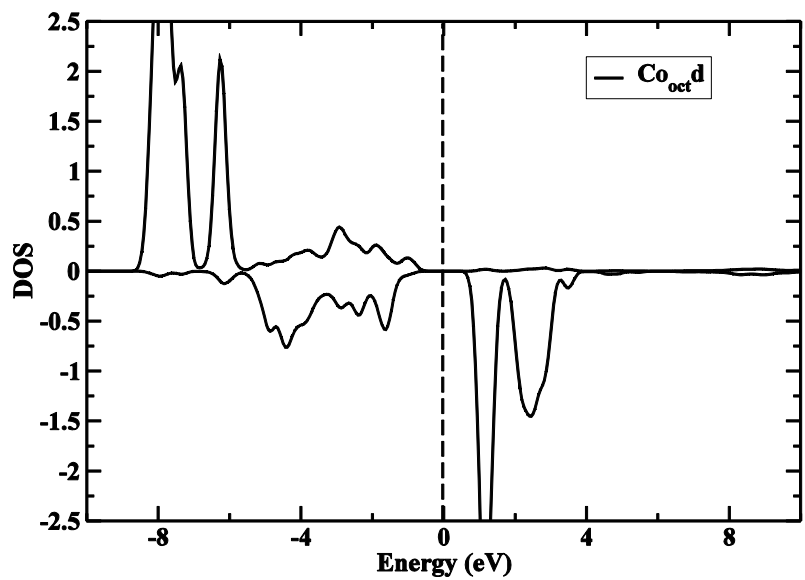

(a)

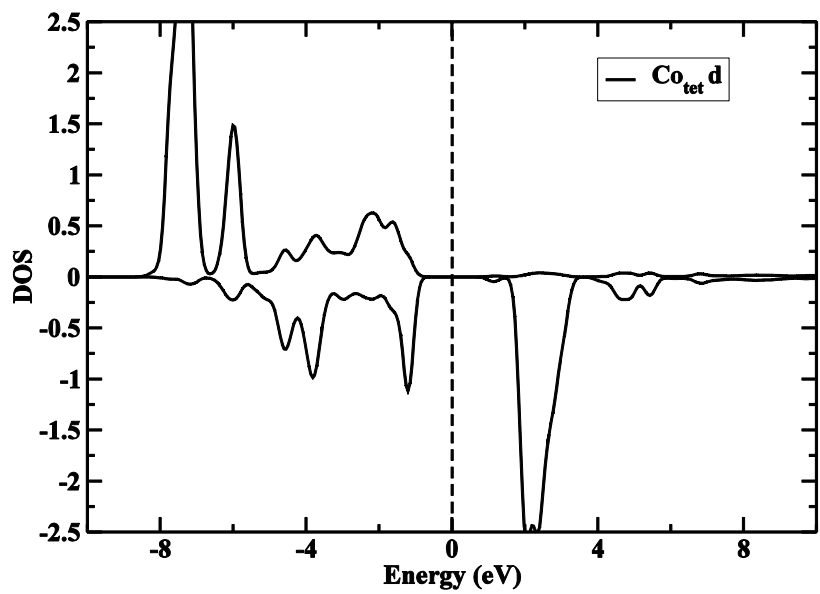

(b)

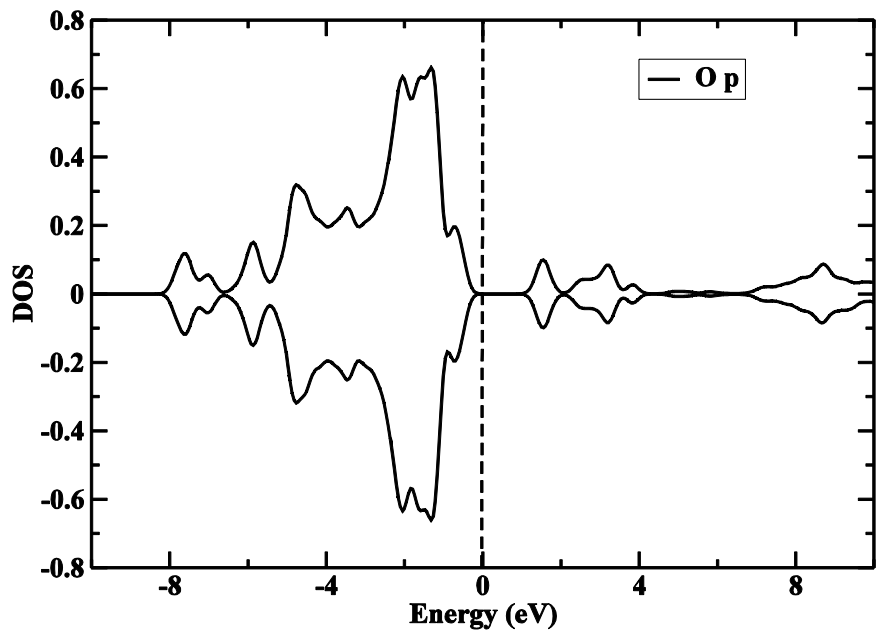

(c)

Fig. 2

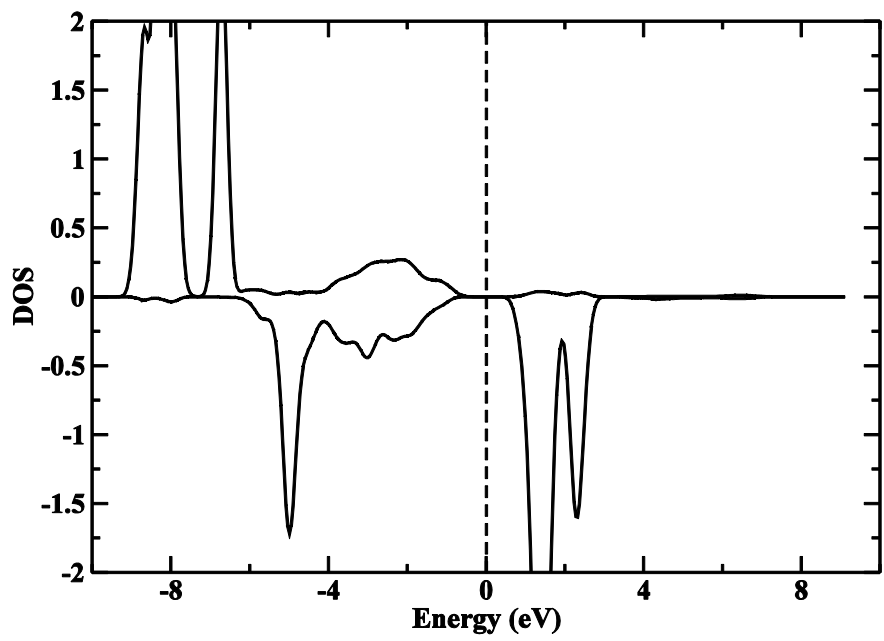

(a)

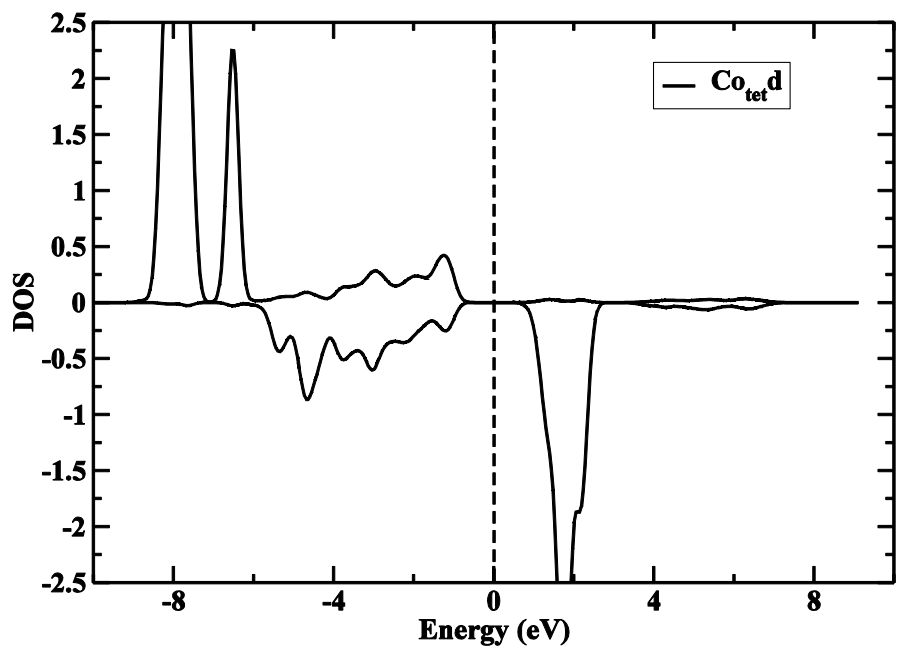

(b)

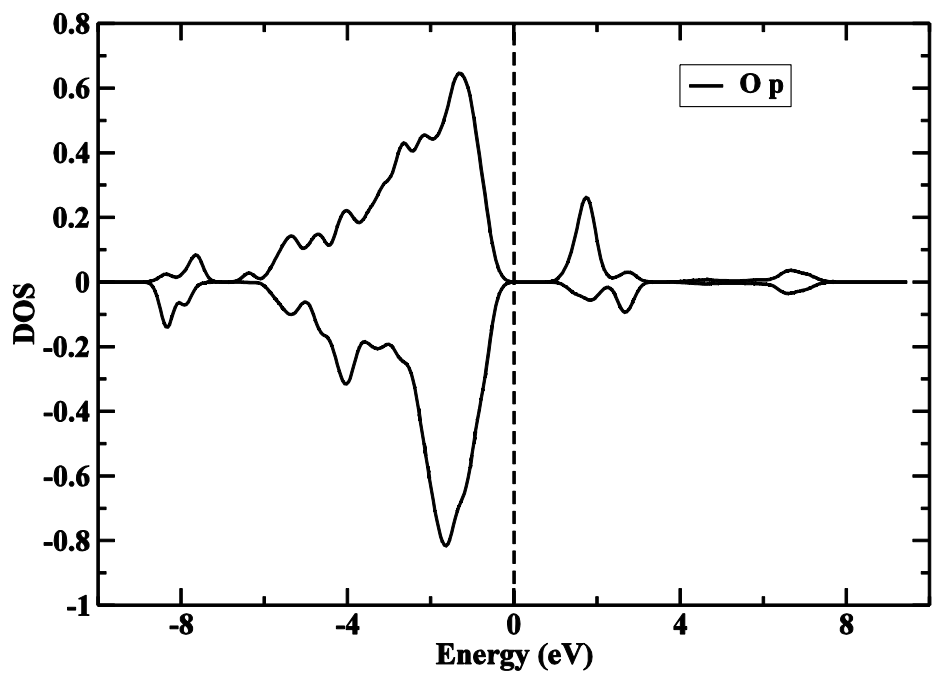

(c)

Fig. 3

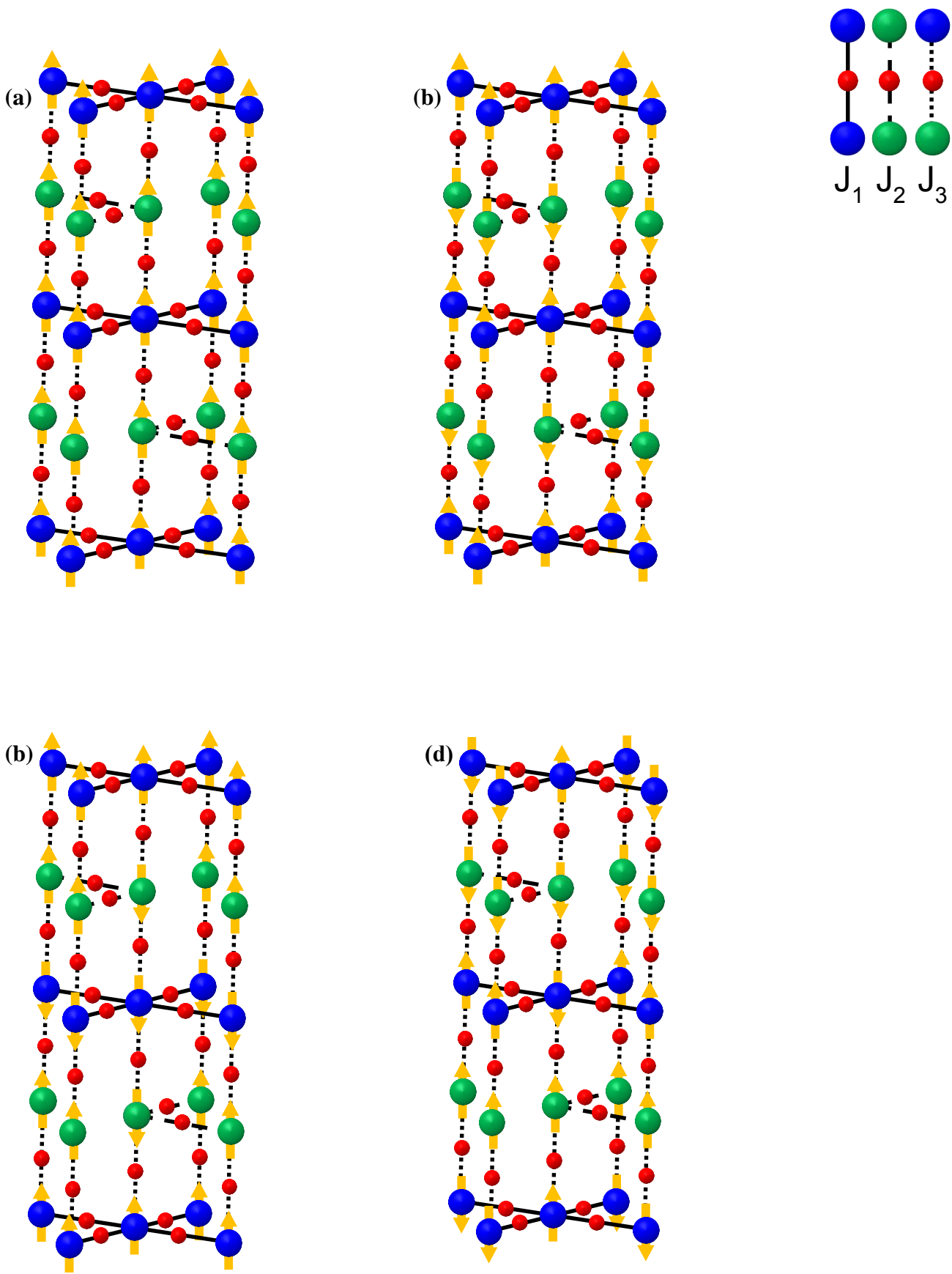

**Fig. 4**

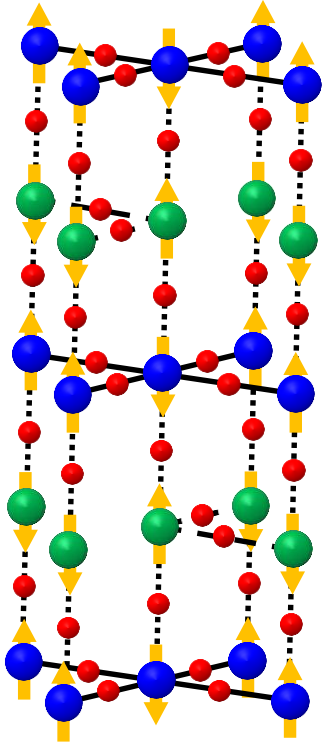

**Fig. 5**

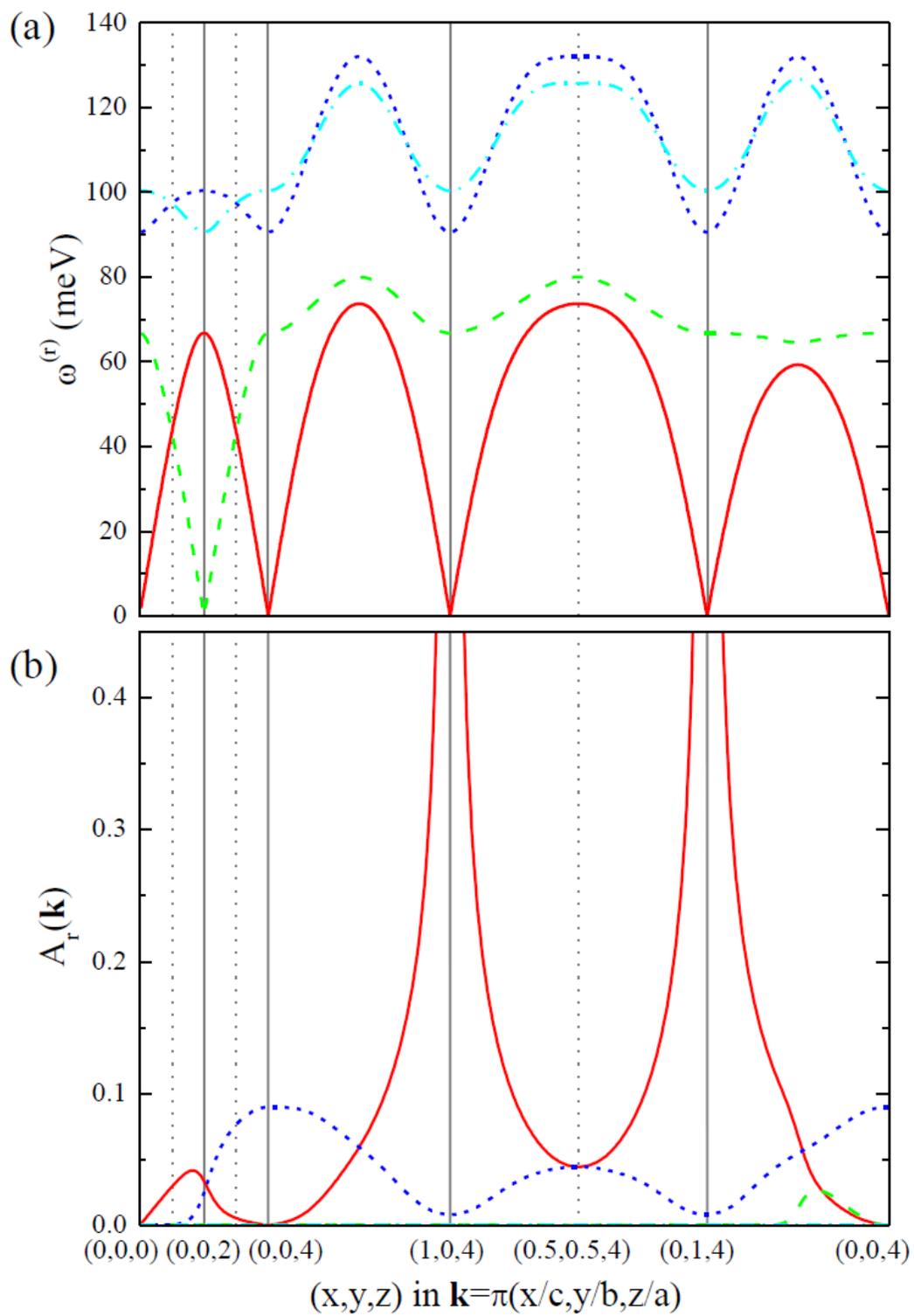

**Fig. 6**